\documentclass[a4paper, 12 pt, fleqn]{article}

\usepackage{bm}
\usepackage{booktabs}
\usepackage{natbib} % serve per \citep \citet
\usepackage{graphicx}

\usepackage{authblk}
\author[1]{B. J. Devenish\footnote{Corresponding author: B. J. Devenish, {\it ben.devenish@metoffice.gov.uk}}}
\author[2]{M. Cerminara}
\title{A Lagrangian stochastic model of a volcanic eruption column}
\affil[1]{\small Met Office, FitzRoy Road, Exeter, EX1 3PB, UK}
\affil[2]{Istituto Nazionale di Geofisica e Vulcanologia, Sezione di Pisa, Italy}

\begin{document}
\maketitle
\begin{abstract}
We develop a Lagrangian stochastic model (LSM) of a volcanic plume in which the mean flow
is provided by an integral plume model of the eruption column and fluctuations in the
vertical velocity are modelled by a suitably constructed stochastic differential equation.
The LSM is applied to the two eruptions considered by \citet{Costa2016} for the
volcanic-plume intercomparison study. Vertical profiles of the mass concentration computed from the 
LSM are compared with equivalent results from a large-eddy simulation (LES) for the case of no ambient wind. 
The LSM captures the order of magnitude of the LES mass concentrations and some aspects of their profiles. 
In contrast with a standard integral plume model, i.e. without fluctuations, the mass concentration computed
from the LSM decays (to zero) towards the top of the plume which is consistent with the LES plumes. 
In the lower part of the plume, we show that the presence of ash leads to a peak in the mass concentration
at the level at which there is a transition from a negatively buoyant jet to a positively buoyant plume.
The model can also account for the ambient wind and moisture.
\end{abstract}

%% ------------------------------------------------------------------------ %%
%
%  TEXT
%
%% ------------------------------------------------------------------------ %%

%%% Suggested section heads:
% \section{Introduction}
% 
% The main text should start with an introduction. Except for short
% manuscripts (such as comments and replies), the text should be divided
% into sections, each with its own heading. 

% Headings should be sentence fragments and do not begin with a
% lowercase letter or number. Examples of good headings are:

% \section{Materials and Methods}
% Here is text on Materials and Methods.
%
% \subsection{A descriptive heading about methods}
% More about Methods.
% 
% \section{Data} (Or section title might be a descriptive heading about data)
% 
% \section{Results} (Or section title might be a descriptive heading about the
% results)
% 
% \section{Conclusions}

\section{Introduction}

Volcanic plumes represent the most powerful naturally occuring buoyant sources of airborne
contaminants. The spread of volcanic ash downwind of the eruption column presents a 
significant hazard to aviation which motivates the development of mathematical models
that enable its prediction. Models of the eruption column play a role in quantifying 
a volcanic source for a long-range dispersion model. While simple one-dimensional 
integral models of volcanic plumes 
\citep[see e.g.][and references therein]{Woods88,Woods93,Glaze97,Mastin2007,Devenish2013,Costa2016} 
can provide the variation with height of bulk properties of the plume including the
mass concentration, this latter quantity has the unfortunate property that it blows 
up at the top of the plume. The vertical profile of mass concentration produced by an integral
model is then not suitable for initialising a long-range dispersion model. 
One purpose of this letter is to present a model that can provide
a more realistic vertical profile of the mass concentration. 

A second motivation for this study is the development of a model for the explicit
treatment of volcanic plumes within a long-range dispersion model. 
Models of turbulent dispersion often take a Lagrangian form which
provides a natural framework for modelling, for example, dispersion from a point
source which is harder to model with an Eulerian approach. 
Typically, thousands of model particles
are followed through a given flow field and statistics such 
as the mean concentration are calculated from the ensemble
of particles. The resolved part of the flow field would, for
realistic applications, normally be taken from a numerical weather 
prediction model while the unresolved part of the motion
is modelled by means of random increments to the velocity
of the particles. These models, which are known as Lagrangian 
stochastic models (LSMs), can be rigorously formulated
\citep{Thomson87} and have been very successful at reproducing
observations \citep[e.g.][]{Wilson2012}. 

In most realistic dispersion models that are used for
operational purposes, the Lagrangian particles move independently 
of each other through the flow field. There is then an inherent difficulty
in modelling a coherent process such as a volcanic plume
using single-particle LSMs: the motion of individual particles
or fluid elements depends on the buoyancy of all the fluid elements.
Moreover, there is nothing to constrain two neighbouring model particles to be moving 
upwards with similar velocities.

Several authors have attempted to model simple Boussinesq plumes using a 
Lagrangian approach \citep[e.g.][]{Luhar92,Anfossi93,Weil94,Heinz99,Alessandrini2013,Marro2014}. 
In particular we consider the approach developed by \citet{Webster2002} and \citet{Bisignano2015}
in which the mean flow is calculated from an integral plume model and the fluctuations
are calculated using a suitably formulated stochastic differential equation (sde). 
Here we extend this approach to the modelling of volcanic plumes. 

In the next section we present the LSM which is formulated for a realistic atmosphere with 
non-uniform stability, ambient wind and moisture. In section \ref{numerical} we consider numerical solutions
of the LSM for various different cases: in particular we compare solutions of the LSM with an equivalent
large-eddy simulation (LES) in the case of no ambient wind. 

\section{Lagrangian Stochastic Model} 

The model of \citet{Devenish2013} is used to provide the mean flow. The governing equations take the form
\begin{eqnarray}
\frac{dQ_m}{dt} &=& E \overline{v}_p \label{eq:mass_flux} \\
\frac{dM_z}{dt} &=& (\rho_a - \rho_p) g \pi b^2 \overline{v}_p \label{eq:vertical_momentum} \\
\frac{dM_i}{dt} &=& -Q_m \frac{d U_i}{d t} \qquad i = x,y \label{eq:horizontal_momentum} \\
\frac{dH}{dt}   &=& \left( (1 - q_{va}) c_{pd} + q_{va} c_{pv} \right) T_a E \overline{v}_p - g \rho_a \pi b^2 \overline{v}_p \overline{w}_p \nonumber \\
                 &&    + \left[ L_{v0} - 273 (c_{pv} - c_{pl}) \right] \frac{dQ_l}{dt}  \label{eq:enthalpy_flux} \\
\frac{dQ_t}{dt} &=& E q_{va} \overline{v}_p \label{eq:moisture_flux} 
\end{eqnarray}
where, at time $t$, $Q_m = \rho_p \pi b^2 \overline{v}_p$ is the mass flux;
$Q_t = n_t Q_m$ is the total moisture flux (water vapour and liquid water; the model contains no ice); 
$Q_l = n_l Q_m$ is the flux of liquid water; 
$M_i = (u_{pi} - U_i) Q_m$ ($i=x,y$) are the horizontal components of the momentum flux;
$M_z = \overline{w}_p Q_m$ is the vertical component of the momentum flux; and $H = c_{pp} T_p Q_m$ is the enthalpy flux. 
In equations (\ref{eq:mass_flux})--(\ref{eq:moisture_flux}) and the expressions for the fluxes $\overline{v}_p = \sqrt{u_{px}^2 + u_{py}^2 + \overline{w}_p^2}$
is the total velocity in which $u_{pi}$ ($i=x,y$) are the horizontal components of the plume velocity and $\overline{w}_p$ is the vertical component
of the plume velocity; $n$ is a mass fraction and the subscripts $l$, $v$ and $t$ refer to liquid water, water vapour and the total moisture content respectively 
and we have $n_t = n_v + n_l$; 
$c_{pp}$ is the bulk specific heat capacity of the plume (to be defined below); $b$ is the plume radius; $g$ is the acceleration due to gravity; 
$q_v$ is the humidity; $\rho$ is the density and $T$ is the plume temperature; a subscript $a$ refers to ambient whereas a subscript $p$ refers to plume. The horizontal coordinates
are $x$ and $y$ and $z$ indicates the vertical coordinate. In equation (\ref{eq:horizontal_momentum}) the components of the ambient wind speed
are indicated by $U_i$ ($i=x,y$). In equation (\ref{eq:enthalpy_flux}) $c_{pd}$, $c_{pv}$ and $c_{pl}$ are the specific heat capacities at constant pressure of the
dry air, water vapour and liquid water respectively; $L_{v0}$ is the latent heat of vaporisation at 0$^{\mbox{\footnotesize{o}}}$ C.

The entrainment, $E$, is determined following \citet{Devenish2013} as are phase changes between water vapour and liquid water. 
Similarly, the plume density, bulk gas constant and bulk specific heat capacity are also calculated following \citet{Devenish2013}. 
The evolution of the mass fraction of gas is determined assuming that there is no fall out of material (both ash and liquid water) 
during the evolution of the plume (see \citet{Devenish2013} for more details).

In the construction of the model, it is useful to re-write equation (\ref{eq:vertical_momentum}) as an equation
for the vertical velocity:
\begin{equation}
\frac{d\overline{w}_p}{dt} = \frac{g (\rho_a - \rho_p)}{\rho_p} - \frac{\overline{v}_p \overline{w}_p E}{Q_m}. 
\label{eq:mean_w}
\end{equation}
Equation (\ref{eq:mean_w}) makes clear that the evolution of $\overline{w}_p$ depends on the local buoyancy of the plume and 
that entrainment produces a deceleration of the plume. The sde. for
the fluctuating vertical velocity, $w_p'$, is constructed from an analogous equation to equation (\ref{eq:mean_w}) coupled with an LSM 
for $w_p'$ appropriate for inhomogeneous turbulence. Since we assume there are no fluctuations in $\rho$ (either in the plume or in the environment), the sde for $w_p'$ is given by
\begin{equation}
dw_p' = -\frac{\overline{v}_p w_p' E}{Q_m} dt -\frac{w_p'}{T_L} dt 
           + \frac{1}{2} \left( \frac{1}{w_p} + \frac{w_p'}{\sigma_w^2} \right) d\sigma_w^2 
           + \frac{\sigma_w^2}{w_p \rho_p} \, d \rho_p + \sqrt{C_0 \varepsilon} \, dW
\label{sde:w_prime}
\end{equation}
where $w_p= \overline{w}_p + w_p'$, $T_L$ is the time scale on which $w_p'$ changes, $\sigma_w^2$ is the vertical-velocity variance, 
$\varepsilon$ is the mean kinetic energy dissipation rate, $dW$ is the increment of a Wiener process and $C_0$ is the constant of 
proportionality in the second-order Lagrangian velocity structure function which typically has a value in the range $5-7$ for 
homogeneous isotropic turbulence \citep[e.g.][]{Yeung2002}; we choose $C_0=6$. The first term on the right-hand side (rhs) of 
equation (\ref{sde:w_prime}) represents entrainment-related turbulence and is motivated by the form of equation (\ref{eq:mean_w})
and consistency with \citet{Bisignano2015}. The last four terms on the rhs of equation (\ref{sde:w_prime}) are those of 
\citet{Thomson87}'s LSM for inhomogeneous turbulence
(as would be the case for a one-point Gaussian joint velocity-density distribution). Note that the penultimate term on 
the rhs is often neglected \citep{Stohl99} but, as will be shown below, can be significant.

It remains to specify the forms of $\sigma_w^2$, $T_L$ and $\varepsilon$ which are all functions of $z$. 
We expect $\sigma_w$ to scale with $\overline{w}$ and $T_L$ to be related to the appropriate mean quantities 
in the problem. Hence, following \citet{Bisignano2015}, we choose $\sigma_w = \alpha |\overline{w}_p|$, 
and $T_L = b/|\overline{w}_p|$. At the vent, $T_L$ is defined by the source radius and the exit velocity which
is consistent with the eddy decorrelation time scale identified by \citet{Cerminara2016}. Since 
\[
T_L = \frac{2 \sigma_w^2}{C_0 \varepsilon},
\]
\citep[e.g.][p.486]{Pope2000} it follows that 
\[
\varepsilon = \frac{2 \alpha^2 \overline{w}_p^2 |\overline{w}_p|}{C_0 b}.
\]

\section{Numerical Solution of the LSM} \label{numerical} 

The LSM presented in the previous section i.e. equations 
(\ref{eq:mass_flux}), (\ref{eq:horizontal_momentum}) -- (\ref{eq:moisture_flux}), (\ref{eq:mean_w}) and (\ref{sde:w_prime}) 
are solved simultaneously using an Euler-Maruyama method. The fluctuating velocity component is initially drawn
from a Gaussian distribution with mean zero and variance $\sigma_w^2$. The results are
computed following 100,000 independently moving particles. Integration is terminated for each 
particle when $\overline{w}_p$ becomes zero. The mass concentration per unit length, $C$, is calculated
according to 
\[
C(z) = \frac{Q_{m0}}{\Delta z \, N_p} \sum_{N_t} (\# \, \mbox{particles in each box} ) \, \Delta t
\] 
where $\Delta z$ is the depth of each box, $N_p$ is the number of particles, $N_t$ is the number
of time steps and $\Delta t$ is the time step (or residence time). 

Results are presented for two eruptions: the weak and strong eruptions considered
by \citet{Costa2016} with and without the ambient wind. The weak eruption is the 26 January 2011 
Shinmoe-dake eruption in Japan that produced a plume that reached about 8 km above sea level 
\citep{Hashimoto2012,Kozono2013,Suzuki2013}. The strong eruption is based on the climactic 
phase of the Pinatubo eruption, Philippines, on 15 June 1991, during which the eruption column 
reached about 39 km above sea level \citep{Koyaguchi93,Holasek96,Costa2013}.
The source mass fluxes for each case are 1.5 Gg s$^{-1}$ and 1.5 Tg s$^{-1}$ respectively. 
The same initial conditions and profiles of the ambient quantities as used by \citet{Costa2016} are
also used here.

The results of the LSM are compared with corresponding results from a recent LES \citep{Cerminara2016} for the
case of no ambient wind. 
The LES plume region is defined as the subdomain where the mass fraction of a tracer is larger than 0.1\% of its
initial value. The maximum rise height of the LES plumes is determined from the mass 
flux: it is the level at which the mass flux falls below 1\% of its value at the vent \citep[see][for more details]{Cerminara2016}. 
The level of neutral buoyancy, $z_{eq}$, is 9.3 km above mean sea level (msl) for the weak eruption and 20.3 km (above msl) 
for the strong eruption \citep{Cerminara2016}. The {\em jet length scale}, $L_M$, is the characteristic height at which the initial 
momentum-dominated jet becomes a buoyancy dominated plume. For a volcanic plume, $L_M$ is defined as \citep[][Section 3.6]{Cerminara_thesis}
\[
L_M = \frac{L_0}{(2 \alpha \mbox{Ri} )^{1/2}}
\]
where
\[
L_0 = \frac{Q_m}{\sqrt{\pi \rho_a M_z}} = b \sqrt{\frac{\rho_p}{\rho_a}} \sqrt{\frac{\overline{v}_p}{\overline{w}_p}}
\]
\[
\mbox{Ri} = \frac{\phi \, g Q_m^3}{\sqrt{\pi \rho_a M_z^5}} = \frac{\phi g b}{\overline{w}_p^2}  
\sqrt{\frac{\rho_p}{\rho_a}} \sqrt{\frac{\overline{v}_p}{\overline{w}_p}}
\]
and $\phi = h_p/h_a - 1$ where $h_p$ and $h_a$ are the specific enthalpies of the plume and the environment respectively. 
The plume properties and those of the environment are evaluated at the vent level. In the Boussinesq limit $L_0$ reduces 
to the vent radius for a vertically rising plume; in practice $\overline{v}_p \approx \overline{w}_p$ for a volcanic plume 
at the vent. In the same limit, $L_M$ reduces to the jet length scale defined by \citet{Morton1959}. For forced plumes such 
as volcanic plumes, $L_0 \ll L_M$. For the weak eruption $L_M \approx 0.35$ km and for the strong eruption $L_M \approx 4$ km.
At the vent level the plume is a mixture of three components: 
water, coarse particles and fine particles. For the weak eruption, the mass fractions are 3 wt.\%, 
48.5 wt\% and 48.5 wt.\% respectively; coarse particles have a diameter of 1 mm and fine particles a diameter of 62.5 $\mu$m.
For the strong eruption the mass fractions are 5 wt.\%, 47.5 wt\% and 47.5 wt.\% respectively; coarse particles have a diameter 
of 500 $\mu$m and fine particles a diameter of 16 $\mu$m. The LES results to be presented below are for the upwardly rising 
core of the plume. 

\begin{figure}[ht]
 \centering
 \includegraphics[width=2.5in]{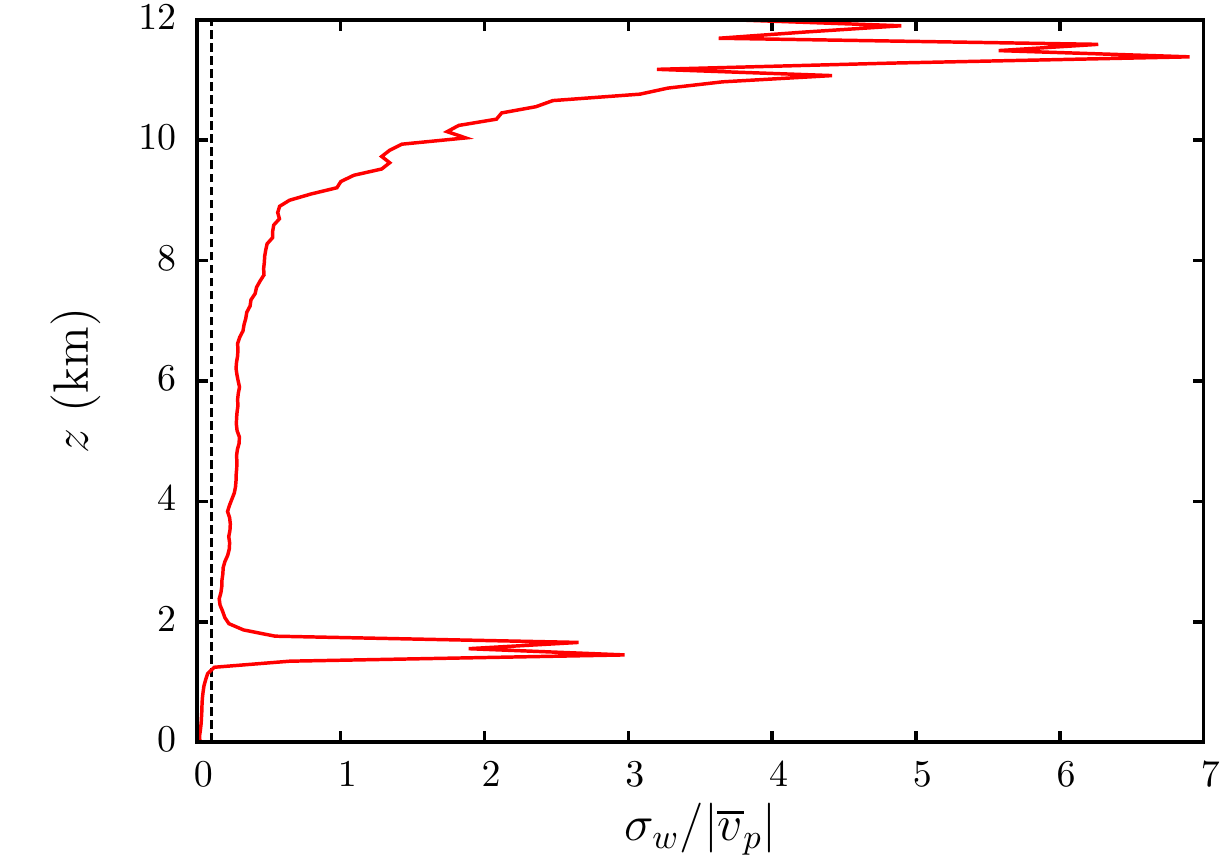}
 \quad
 \includegraphics[width=2.5in]{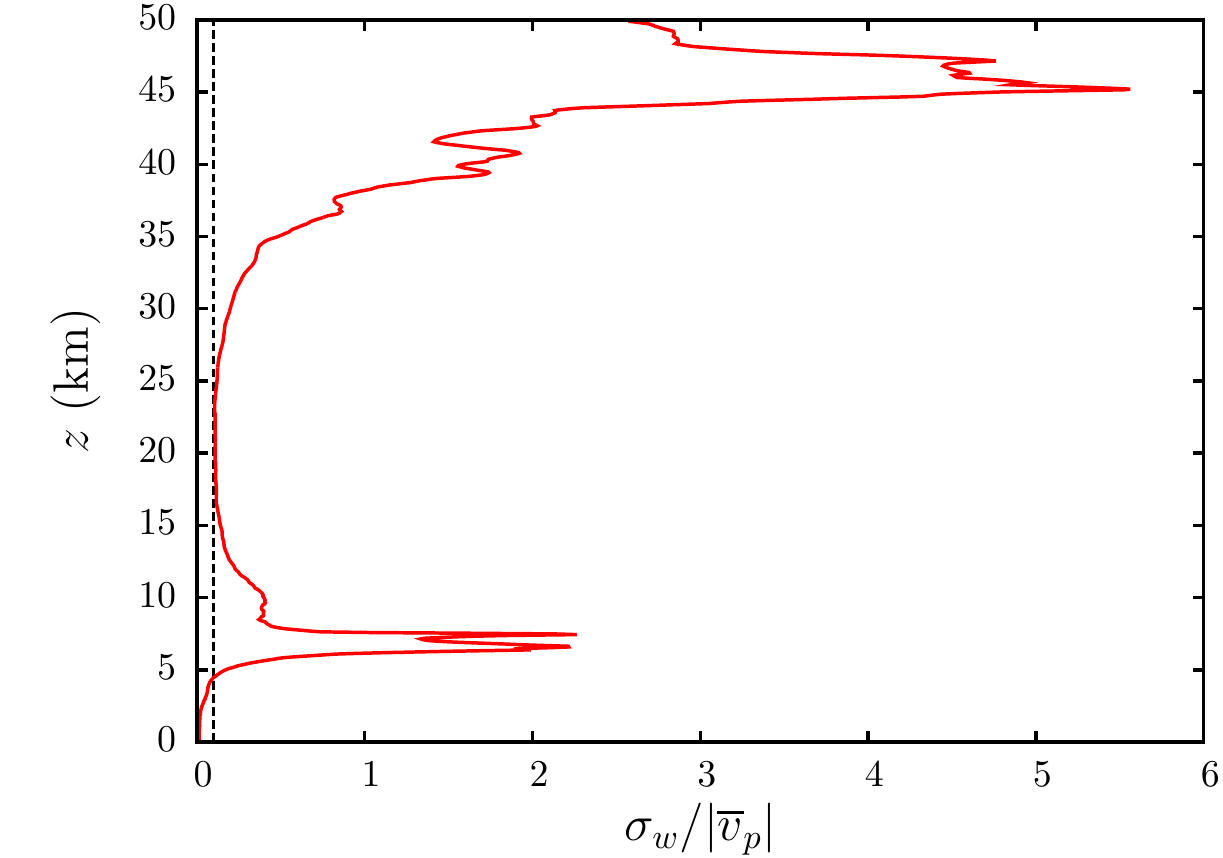}
 \caption{The variation with height of $\sigma_w/|\overline{v}_p|$ (red) for the LES plume for the (a) weak
and (b) strong eruptions. In both figures the dashed black line represents the value $\sigma_w/|\overline{v}_p| = 0.1$.}
 \label{fig:les_w}
\end{figure}
Figure \ref{fig:les_w} shows the variation with height of the ratio $\sigma_w/|\overline{v}_p|$ along the centreline of the LES plumes. (Note that
$\overline{v}_p \approx \overline{w}_p$ over most of the depth of the plume except at heights of order $L_M$ and at the plume top where $\overline{w}_p$ becomes zero.)
It shows that the assumption that $\sigma_w = \alpha \overline{w}_p$ where $\alpha = 0.1$ is reasonable in the positively buoyant part of the plume i.e. the region 
between $L_M$ and $z_{eq}$. 

\begin{figure}[h]
 \centering
 \includegraphics[width=2.5in]{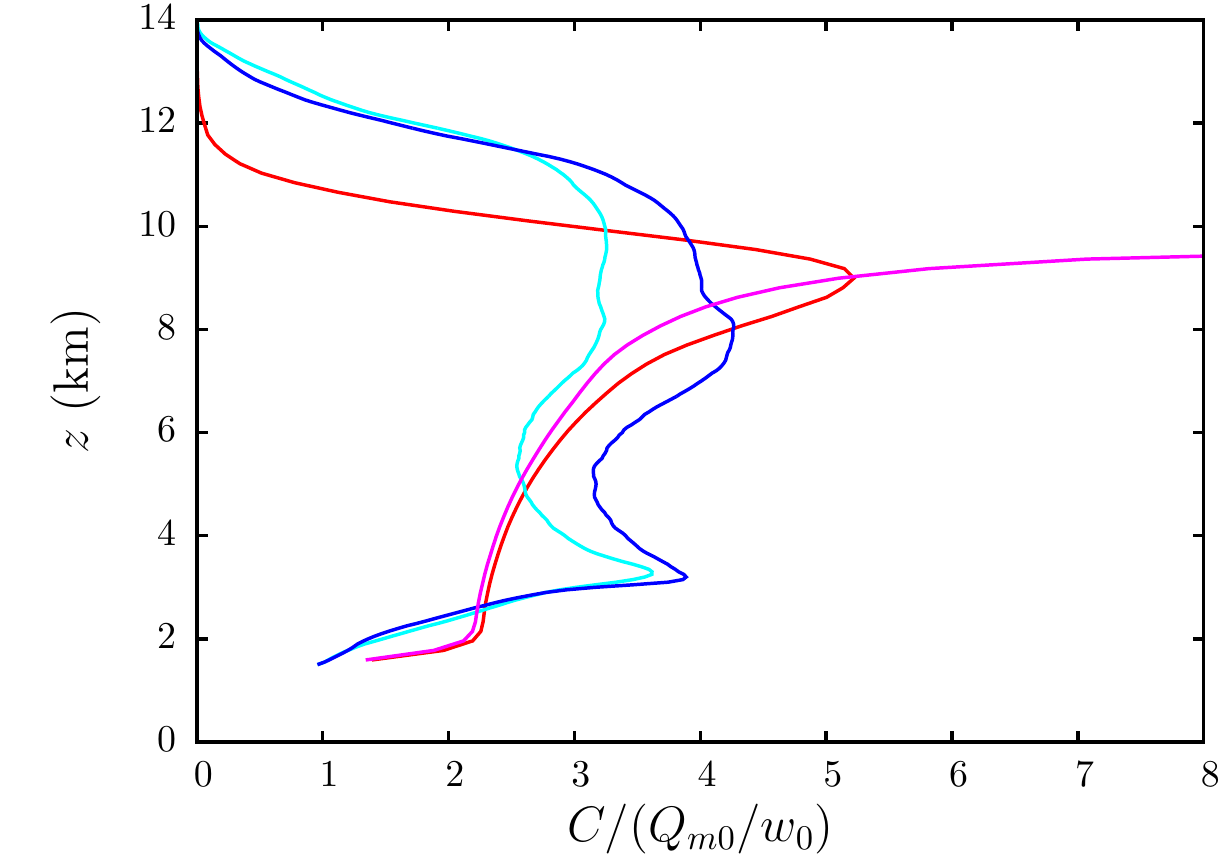}
 \quad
 \includegraphics[width=2.5in]{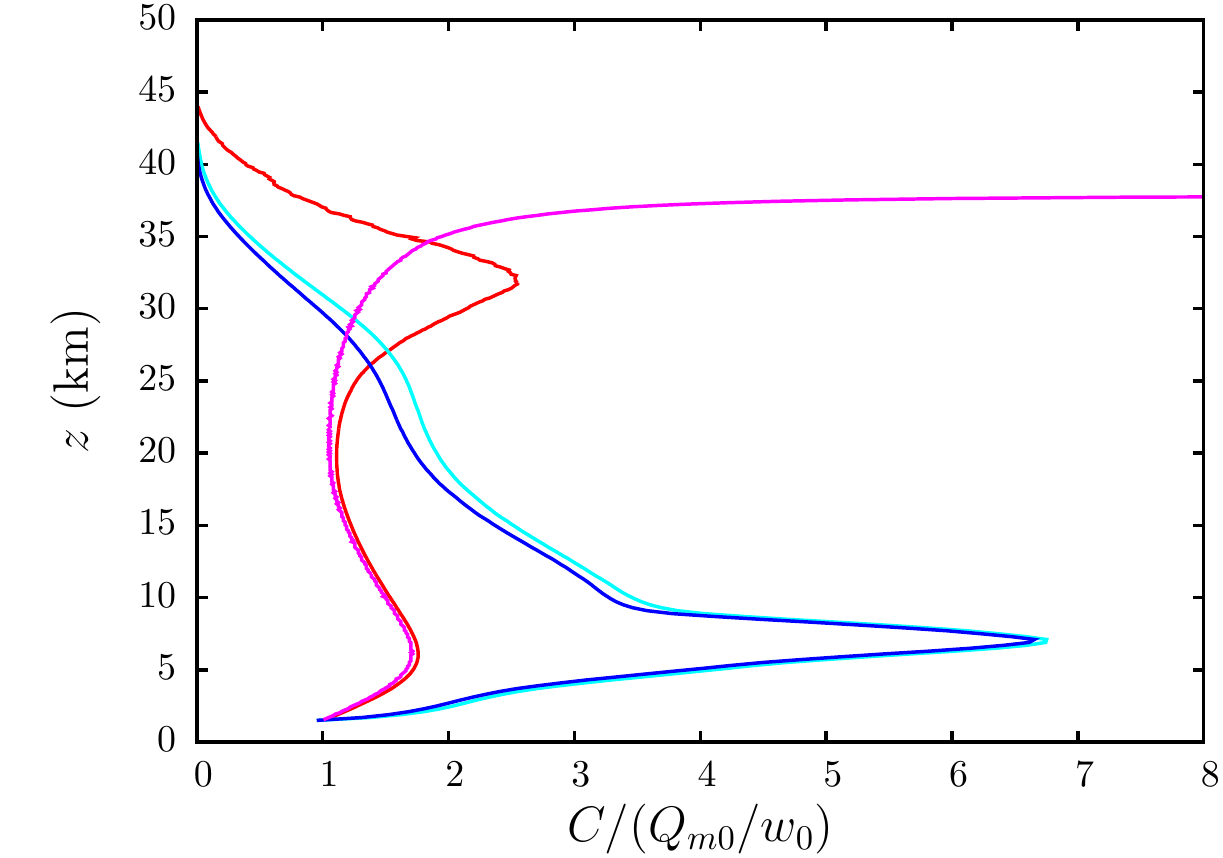}
 \caption{The variation with height of the normalised concentration for the (a) weak and (b) strong eruptions. In each figure the red line represents the LSM, 
the magenta line the LSM with no turbulence, the cyan line the LES with the coarse mass fraction and the blue line the LES with the fine mass fraction. 
In the normalisation of the concentration, the subscript `0' represents the initial value at the vent.}
 \label{fig:concentration_les}
\end{figure}
The vertical profile of the mass concentration in the LSM is shown in Figure \ref{fig:concentration_les} for the case of no ambient wind both with and without turbulence. The latter case 
amounts to solving the plume equations alone i.e. equations (\ref{eq:mass_flux}) -- (\ref{eq:moisture_flux}) such that the concentration is given by $\pi b^2 \rho_s$. 
Figure \ref{fig:concentration_les} clearly shows that turbulent fluctuations have very little effect on the concentration profile in the lower part of the plume but that
without turbulent fluctuations the mass concentration increases very rapidly towards the top of the plume. This behaviour is explained by the fact that as $\overline{w}_p \rightarrow 0$
at the top of the plume, $b \rightarrow \infty$ (since $Q_m$ grows monotonically with height). In the case with turbulent fluctuations, particles have different values of $z$ 
when $\overline{w}_p=0$ and this gives rise to the smooth peak which decays to zero at the top of the plume. 
 
Figure \ref{fig:concentration_les} also shows the vertical profiles of the mass concentration for the LES plume (i.e. $\int_{S(z)} \rho_p(x,y,z) \, dx \, dy$
where $S(z)$ is a horizontal slice at height $z$ of the LES plume region with positive vertical velocity).  
It can be seen that the maximum rise height of the LSM plume is broadly consistent with that of the LES plume. The LSM plume is better at capturing the
order of magnitude of the concentration and the vertical structure of the LES plume for the weak eruption than the strong eruption. This is also evident in the total 
mass within the plume (i.e. $\int C(z) \, dz$) shown in Table \ref{tab:total_mass} for the weak and strong eruptions. The vertical profile of the 
LES plume for the weak eruption shows a distinct lower peak and a broader higher peak. The profile for the strong eruption is dominated by the lower peak. 
The equivalent LSM plumes show pronounced upper peaks but only incipient lower peaks especially for the weak eruption. For both the weak and strong eruptions, the
height of the upper peak in the LSM plume is consistent with the level of maximum radial spread found by \citet{Suzuki2016}. 
The presence of the solid phase in the LES plumes can lead to partial column collapse, particle settling and re-circulation especially at $z \sim L_M$.
This is particularly prevalent in the strong eruption and leads to the formation of the significant lower peak in the concentration profile 
as shown in Figure \ref{fig:concentration_les}b. However, since these processes are absent in the LSM, it does not explain why the LSM plume also 
exhibits a lower peak. 
\begin{table}[h]
\centering
\begin{tabular}{|c || c c c c c c c c c c c c}
  \hline
  Case   & \multicolumn{3}{c}{LSM} & \multicolumn{9}{c}{LES} \\ \hline
         & \multicolumn{3}{c}{}       & \multicolumn{3}{c}{Coarse} & \multicolumn{3}{c}{Fine}   & \multicolumn{3}{c}{Total}  \\ \hline
  Weak   & \multicolumn{3}{c}{317 Gg} & \multicolumn{3}{c}{192 Gg} & \multicolumn{3}{c}{164 Gg} & \multicolumn{3}{c}{356 Gg} \\ \hline
  Strong & \multicolumn{3}{c}{308 Tg} & \multicolumn{3}{c}{195 Tg} & \multicolumn{3}{c}{210 Tg} & \multicolumn{3}{c}{405 Tg} \\ \hline
\end{tabular}
\caption{The total mass in the LSM and LES plumes.}
\label{tab:total_mass}
\end{table}

\begin{figure}[h]
 \centering
 \includegraphics[width=2.5in]{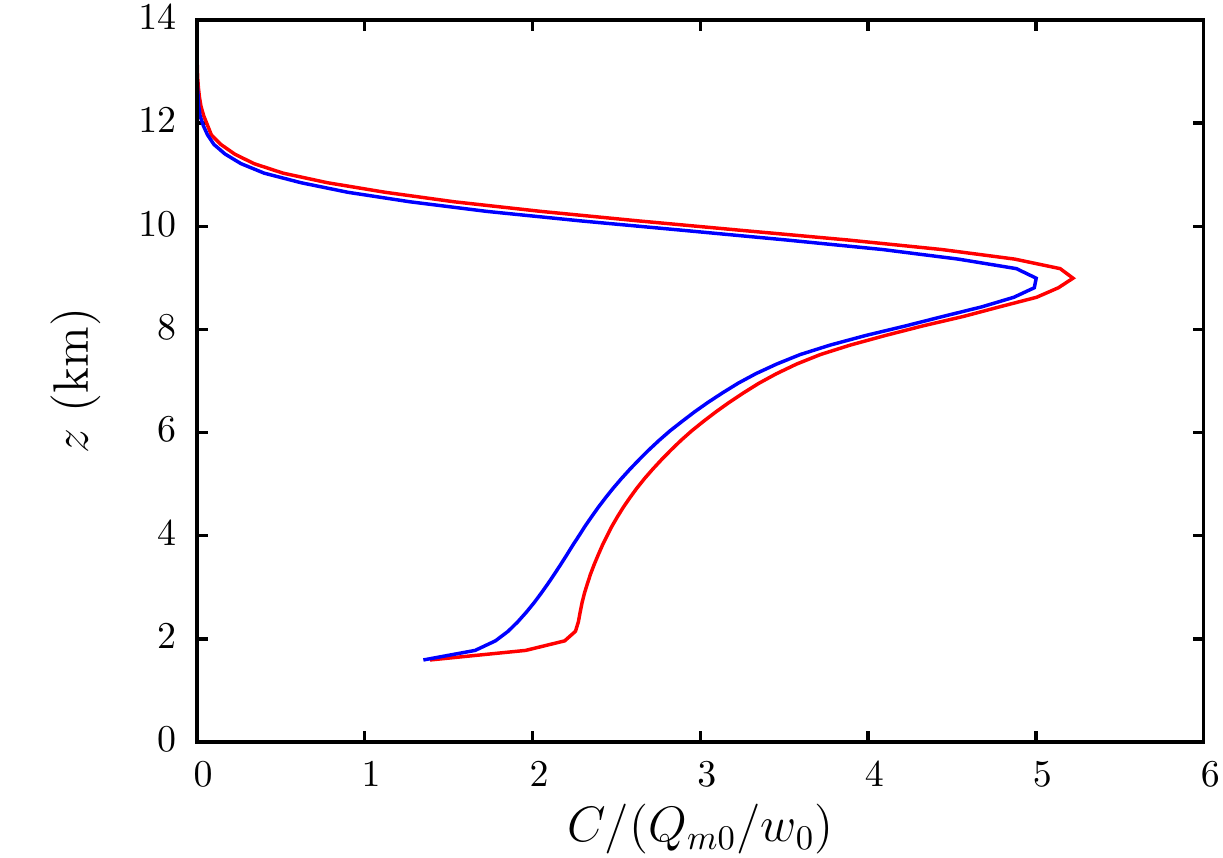}
 \quad
 \includegraphics[width=2.5in]{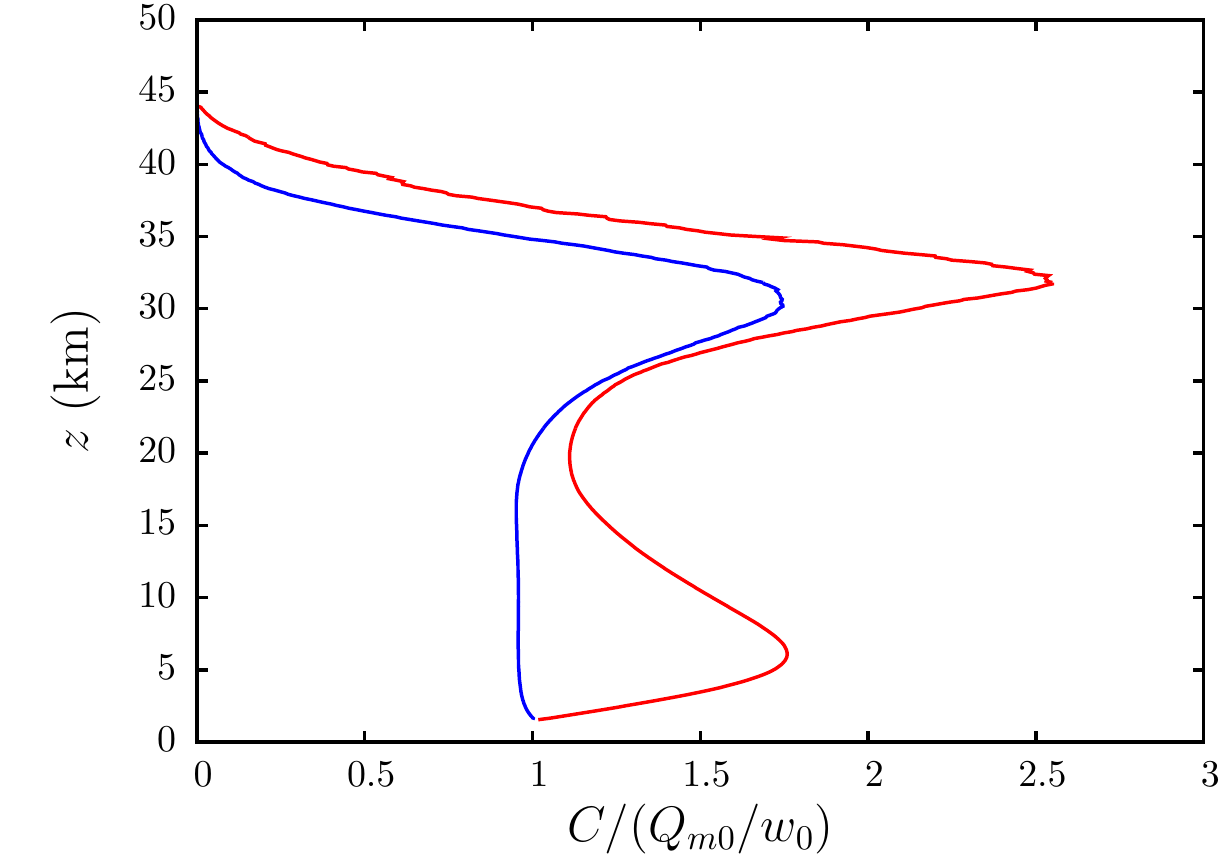}
 \caption{The normalised concentration for the LSM plume: (a) weak and (b) strong eruptions. 
In each figure the red line represents the default case i.e. an ash rich plume (as shown in Figure \ref{fig:concentration_les}) 
and the blue line represents the case of a gas-only plume.}
 \label{fig:concentration_gas_only}
\end{figure}
Figure \ref{fig:concentration_gas_only} shows the LSM plumes with and without ash for both the weak and strong eruptions. In the absence of ash, 
the concentration does not exhibit a lower peak for either the weak or strong eruptions; this is particularly noteworthy for the strong eruption. 
The height of the lower peak occurs, to a good approximation, at a distance $L_M$ above the vent, the height associated with buoyancy reversal 
i.e. a transition from negative to positive buoyancy. In the LES plume this is the height at which partial column collapse  
tends to occur since material above this height is more likely to be carried aloft by the positive buoyancy. In the absence of ash, the plume 
is dominated by buoyancy from the vent upwards and so no transition occurs. 

\begin{figure}[h]
 \centering
 \includegraphics[width=2.5in]{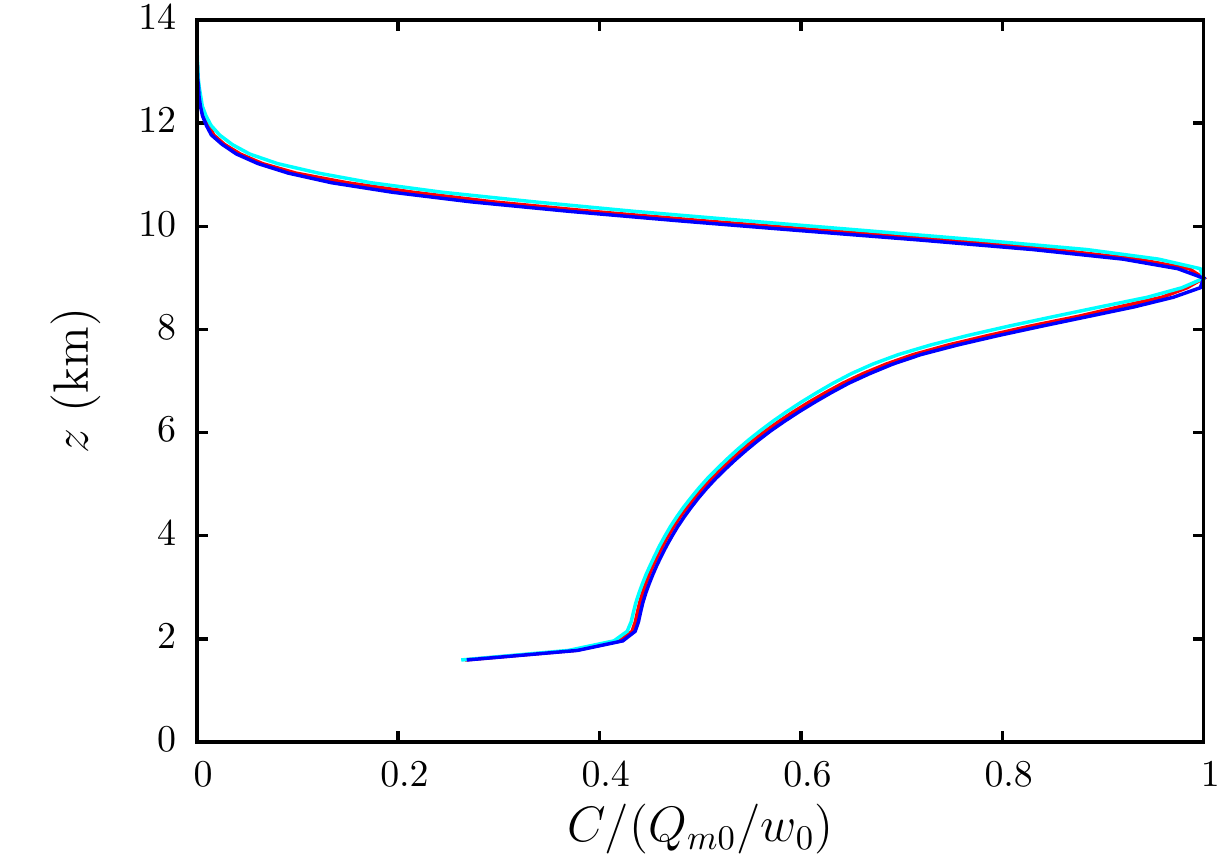}
 \quad
 \includegraphics[width=2.5in]{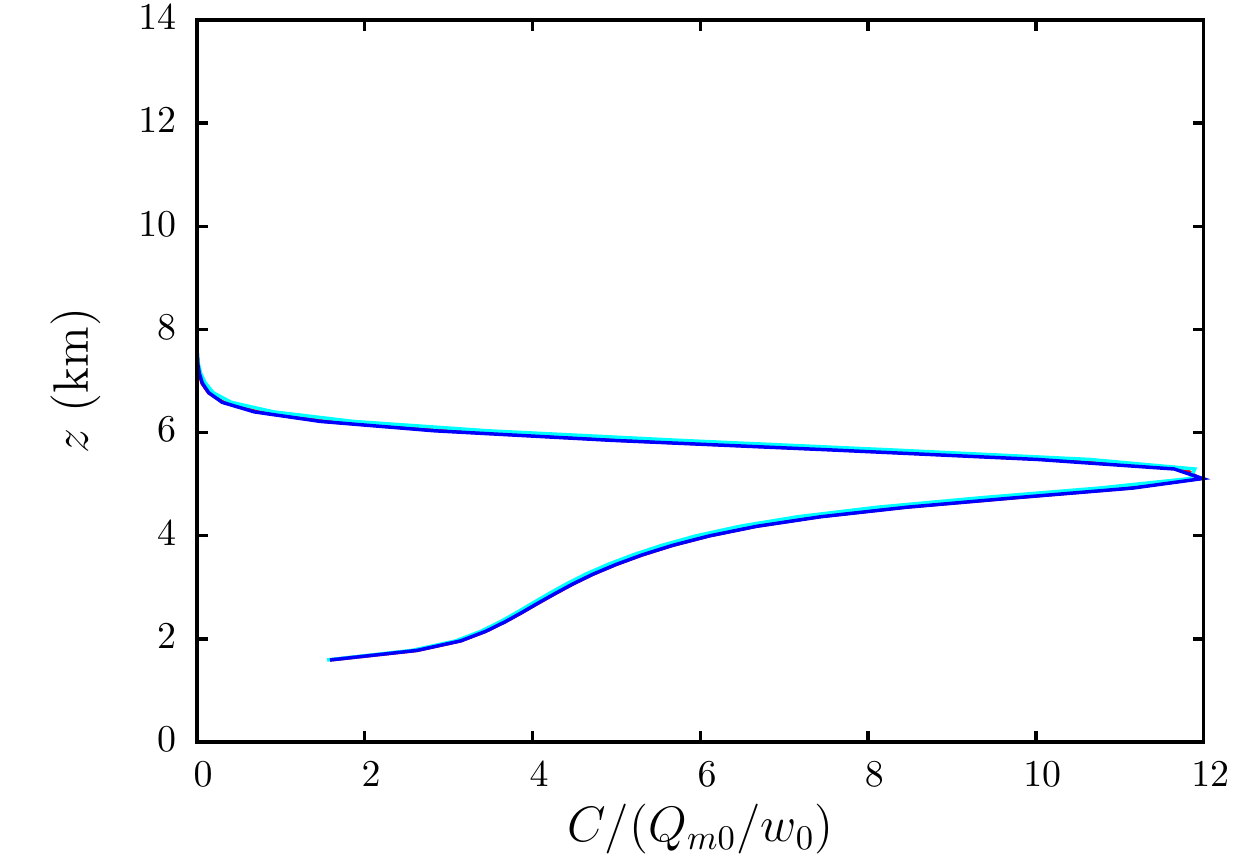}
 \newline
 \includegraphics[width=2.5in]{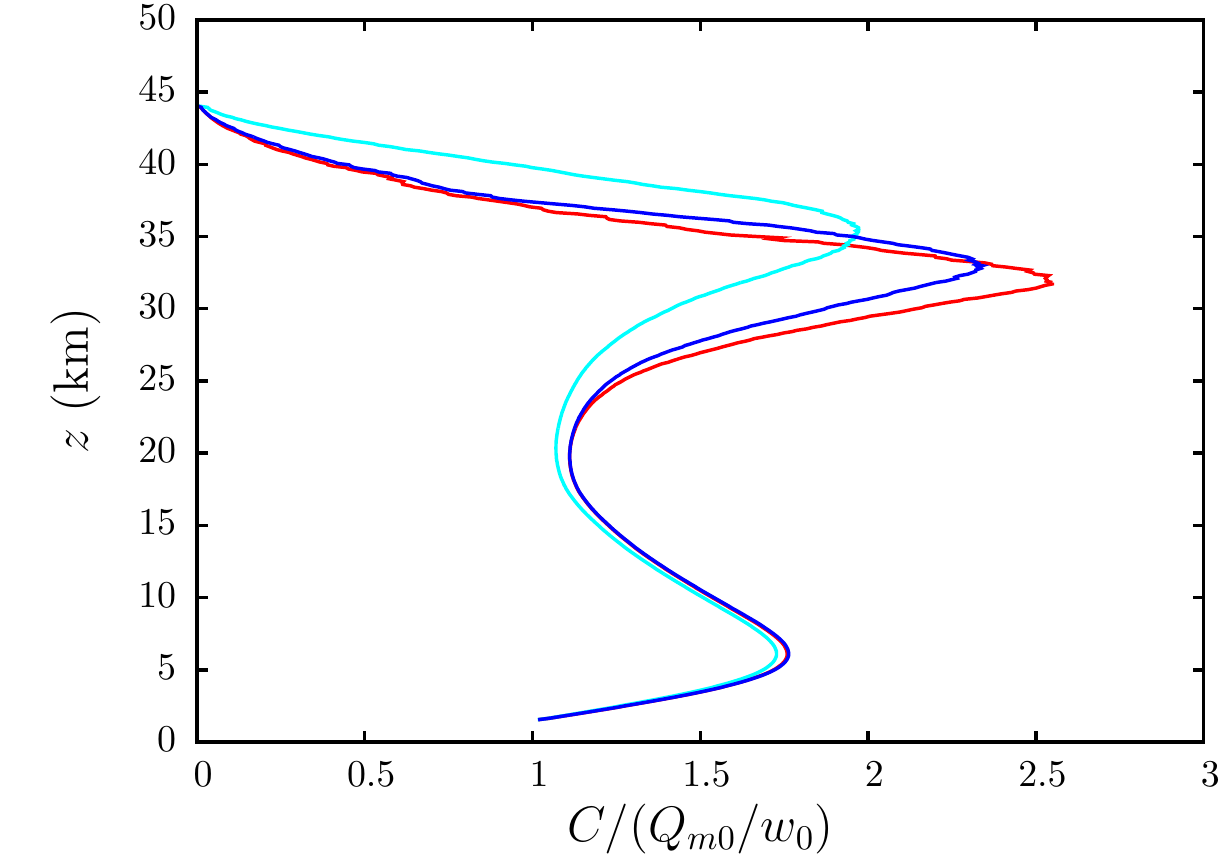}
 \quad
 \includegraphics[width=2.5in]{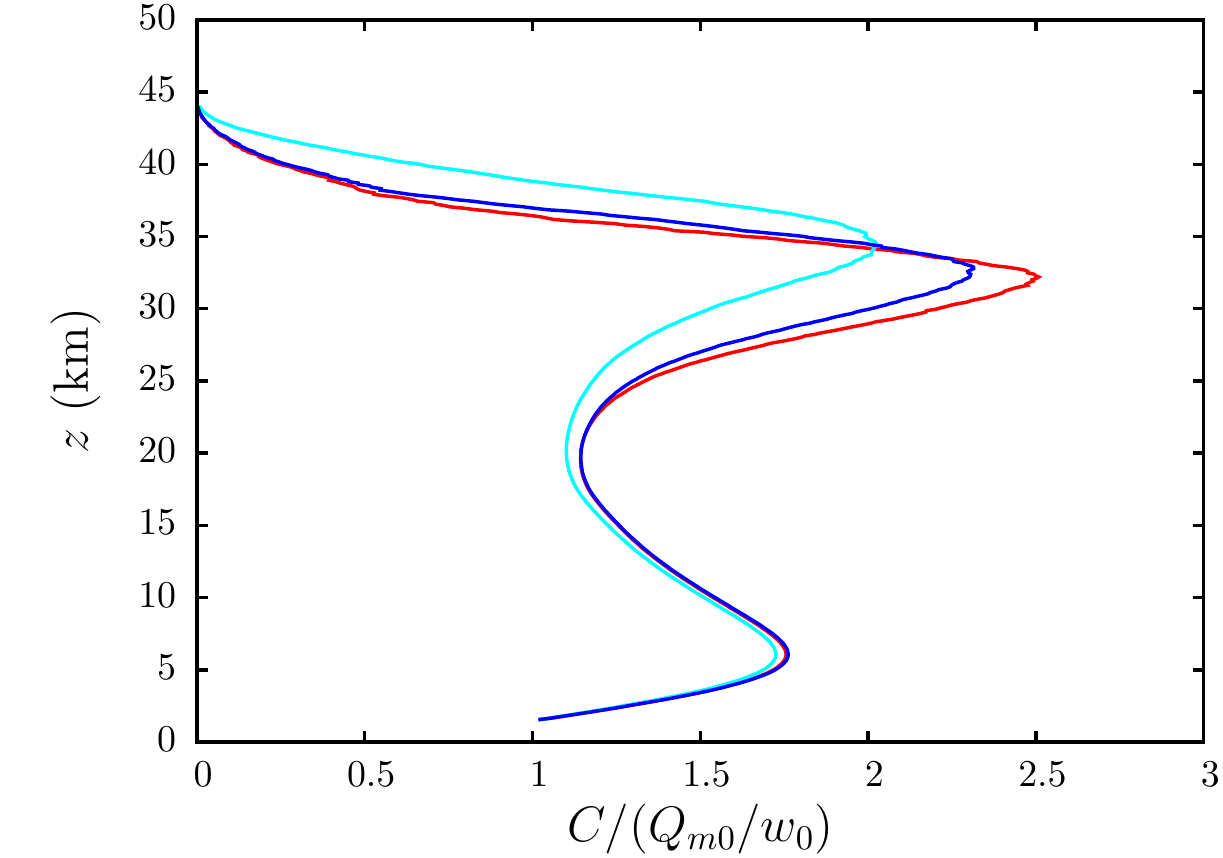}
 \caption{The normalised concentration for the weak (top row) and strong (bottom row) eruptions. The left-hand column is the case without the ambient wind and the
right-hand column is the case with the ambient wind. In each figure the red line is the default case (as shown in Figure \ref{fig:concentration_les}), the blue line
the LSM without moisture and the cyan line the LSM with no density correction.}
 \label{fig:concentration_lsm_only}
\end{figure}
The LSM can be used to assess the importance of a number of different physical processes. In Figure \ref{fig:concentration_lsm_only} we show the effect of 
the ambient wind, moisture and the acceleration due to the change in density with height (as represented by the penultimate term on the rhs of equation 
(\ref{sde:w_prime})). For all these cases, no equivalent LES data is available. It is immediately clear that the ambient wind has a significant effect 
on the rise height of the weak eruption with no appreciable effect on the qualitative morphology of the concentration profile. It is also clear that 
neither moisture nor the density correction have any noticeable impact on this case. For the strong eruption, it is clear that the ambient wind has no effect 
but that the moisture has a small effect on the vertical structure of the plume; the same can be said for the density correction. 

\section{Conclusions}

We have presented an LSM of a volcanic plume in which an integral volcanic-plume model provides the mean flow and a 
suitably constructed sde gives the fluctuating vertical velocity. We compared the mass concentration computed from the LSM 
with data from a corresponding LES for the two eruptions considered in the volcanic-plume intercomparison study 
of \citet{Costa2016}. The LSM captures the order of magnitude of the mass concentration and aspects of its vertical profile.
In qualitative agreement with the LES results, the LSM mass concentration decays to zero at 
the top of the plume. In contrast, the mass concentration computed from a standard integral plume model, i.e. without fluctuations,
blows up at the top of the plume. As with the integral plume models considered in \citet{Costa2016}
the LSM compares relatively well with the weak eruption but not so well with the strong eruption. 
The reasons for this are complex but are likely to include a relatively small aspect ratio and a
relatively small ratio of $z_{eq}$ to $L_M$ compared with the weak eruption. 
We showed that the presence of ash alone is sufficient to produce a peak in the mass concentration at $z \sim L_M$, the height
at which there is a transition from a negatively buoyant jet to a positively buoyant plume. Although for the LES plume the magnitude of this peak 
is likely to be determined by the complex flow features in the LES, their absence in the LSM and the standard integral plume model suggests that $L_M$
plays an important role in the LES plume as well. 

The LSM developed here is designed to be used with realistic meteorological profiles including the ambient wind and moisture. An analysis of the 
effect of ambient conditions on the two eruption columns showed that, as expected, the ambient wind affected the weak eruption but not the strong
eruption. In contrast, ambient moisture had a small effect on the strong eruption but almost no effect on the weak eruption. 
The LSM can be used to provide a vertical distribution of material for use in an operational dispersion model where the eruption 
column is often modelled as a uniform passive line source. Furthermore, it could be incorporated into a Lagrangian dispersion model 
to provide a dynamic source. 

The LES data can be accessed at the repository:\\
vmsg.pi.ingv.it/uploads/files/publications/LES\_data.zip

\end{document}